\documentclass[twocolumn]{aastex63}
\usepackage{placeins}
\usepackage{comment}
\usepackage{amsmath}
\pdfoutput=1

\newcommand{\sao}{\affiliation{Smithsonian Astrophysical Observatory, Cambridge, MA, USA}}
\newcommand{\umich}{\affiliation{University of Michigan, Ann Arbor, MI, USA}}
\newcommand{\ucb}{\affiliation{Physics Department,University of California, Berkeley, CA, USA}}
\newcommand{\les}{\affiliation{LESIA, Observatoire de Paris, PSL Research University, CNRS, UPMC Université Paris 6, Université Paris-Diderot, Meudon, France}}
\newcommand{\ufi}{\affiliation{Physics and Astronomy Department, University of Florence, Sesto Fiorentino, Italy}}
\newcommand{\ina}{\affiliation{INAF - Osservatorio Astrofisico di Arcetri, Firenze, Italy}}
\newcommand{\ssl}{\affiliation{Space Sciences Laboratory, University of California, Berkeley, CA 94720-7450, USA}}

\newcommand{\imp}{\affiliation{The Blackett Laboratory, Imperial College London, London, SW7 2AZ, UK}}
\newcommand{\que}{\affiliation{School of Physics and Astronomy, Queen Mary University of London, London E1 4NS, UK}}

\newcommand{\orl}{\affiliation{LPC2E, CNRS and University of Orl\'eans, Orl\'eans, France}}

\newcommand{\mine}{\affiliation{School of Physics and Astronomy, University of Minnesota, Minneapolis, MN 55455, USA}}

\newcommand{\nas}{\affiliation{Solar System Exploration Division, NASA/Goddard Space Flight Center, Greenbelt, MD, 20771, USA}}

\newcommand{\lasp}{\affiliation{Laboratory for Atmospheric and Space Physics, University of Colorado, Boulder, CO 80303, USA}}

\received{January 27, 2020}
\accepted{March 2, 2020}
\submitjournal{ApJ}

\shorttitle{Coronal Electron Temperature inferred from the Strahl Electrons}
\shortauthors{Ber\v{c}i\v{c} et al.}

\begin{document}

\title{Coronal Electron Temperature inferred from the Strahl Electrons in the Inner Heliosphere: Parker Solar Probe and Helios observations}

\correspondingauthor{Laura Ber\v{c}i\v{c}}
\email{laura.bercic@obspm.fr}
\author[0000-0002-6075-1813]{Laura Ber\v{c}i\v{c}}\les\ufi
\author[0000-0001-5030-6030]{Davin Larson}\ucb
\author[0000-0002-7287-5098]{Phyllis Whittlesey}\ucb
\author[0000-0001-6172-5062]{Milan Maksimovi\'{c}}\les
\author[0000-0002-6145-436X]{Samuel T. Badman}\ucb
\author[0000-0002-1647-6121]{Simone Landi}\ufi\ina
\author[0000-0002-6276-7771]{Lorenzo Matteini}\les

\author[0000-0002-1989-3596]{Stuart. D. Bale}\ucb\ssl\imp\que
\author[0000-0002-0675-7907]{John W. Bonnell}\ssl
\author[0000-0002-3520-4041]{Anthony W. Case}\sao
\author[0000-0002-4401-0943]{Thierry {Dudok de Wit}}\orl
\author[0000-0003-0420-3633]{Keith Goetz}\mine
\author[0000-0002-6938-0166]{Peter R. Harvey}\ssl
\author[0000-0002-7077-930X]{Justin C. Kasper}\umich\sao
\author[0000-0001-6095-2490]{Kelly E. Korreck}\sao
\author[0000-0002-0396-0547]{Roberto Livi}\ucb
\author[0000-0003-3112-4201]{Robert J. MacDowall}\nas
\author[0000-0003-1191-1558]{David M. Malaspina}\lasp
\author[0000-0002-1573-7457]{Marc Pulupa}\ssl
\author[0000-0002-7728-0085]{Michael L. Stevens}\sao

\begin{abstract}
The shape of the electron velocity distribution function plays an important role in the dynamics of the solar wind acceleration. Electrons are normally modelled with three components, the core, the halo, and the strahl. We investigate how well the fast strahl electrons in the inner heliosphere preserve the information about the coronal electron temperature at their origin. We analysed the data obtained by two missions, Helios spanning the distances between 65 and 215 R$_S$, and Parker Solar Probe (PSP) reaching down to 35 R$_S$ during its first two orbits around the Sun. 
The electron strahl was characterised with two parameters, pitch-angle width (PAW), and the strahl parallel temperature (T$_{s\parallel}$). PSP observations confirm the already reported dependence of strahl PAW on core parallel plasma beta ($\beta_{ec\parallel}$)\citep{Bercic2019}. Most of the strahl measured by PSP appear narrow with PAW reaching down to 30$^o$. The portion of the strahl velocity distribution function aligned with the magnetic field is for the measured energy range well described by a Maxwellian distribution function. T$_{s\parallel}$ was found to be anti-correlated with the solar wind velocity, and independent of radial distance. These observations imply that T$_{s\parallel}$ carries the information about the coronal electron temperature. The obtained values are in agreement with coronal temperatures measured using spectroscopy \citep{David1998}, and the inferred solar wind source regions during the first orbit of PSP agree with the predictions using a PFSS model \citep{Bale2019,Badman2019}.

\end{abstract}



\section{Introduction}
\label{sec:intro}

The solar wind is the constant flux of plasma which leaves the solar corona and expands in our solar system \citep{Parker1958}. It consists of mostly electrons and protons, both exhibiting non-thermal velocity distribution function (VDF) features. Electrons are usually modelled by three components. The lower electron energies are dominated by the core, Maxwellian-like population taking up most of the total electron density.
Electrons with higher energies are either part of the magnetic field-aligned strahl population, or of the halo population present at all pitch angles \citep{Feldman1975a, Pilipp1987b, Maksimovic2005a, stverak2008, Stverak2009a, Tao2016, Wilson2019a, Wilson2019b, Macneil2020}. These models were based on the observations of the solar wind far from the Sun (the closest at 0.3 au), where the solar wind already propagates with a supersonic velocity and where most properties of the pristine coronal plasma have been changed. But how does the electron VDF look like in the solar corona? Does it exhibit high energy tails, or is the excess of the high energy electrons observed in the interplanetary solar wind created during the expansion from purely Maxwellian coronal electrons?

Multi-component distribution functions are used in the kinetic exospheric models of the solar wind initially assuming collisionless evaporation of the solar corona into interplanetary space \citep{Jockers1970, LemaireJosephandScherer1971}. The acceleration of the solar wind in these models is accounted to the solar wind electrons. As their velocities are much higher than the velocities of protons with the same temperature in the solar corona, a portion of electrons manage to escape the Sun and create charge imbalance in the plasma. The imbalance gives rise to an anti-sunward directed electric field, accelerating the heavier solar wind protons. This dynamics produces two main populations in electron VDF. Electrons with energies smaller than the electric potential energy needed to sustain the anti-sunward electric field are bounded to the Sun and present the dense thermal core population. The faster anti-sunward directed electrons, which are able to overcome the potential, escape and form the strahl. The escaping strahl electrons are governed by the magnetic momentum ($\frac{m_e v_\perp^2}{2B}$ = const.) and energy (E$_{kin}$ + E$_{pot}$ = const.) conservation. As they expand into regions with weaker magnetic field they experience focusing \citep{SchwartzMarsch1983}.

Similarly a two-component VDF was obtained by the exospheric models accounting for collisions with Fokker-Planck equation solver using a test particle approach \citep{LieSvendsen1997, pierrard2001}, and by the kinetic simulation of the solar wind accounting for Coulomb collisions statistically \citep{Landi2012b,Landi2014b}. 

These models describe well the formation of the core and the strahl, but they do not explain the formation of the halo. It is possible that the halo is already present in the solar corona, consisting of hot electrons leaking from the dense coronal regions with closed magnetic field loops. Exospheric models assuming an excess of high-energy electrons in the corona were the first models able to self consistently produce fast solar wind reaching velocities above $\sim$ 700 km/s \citep{Maksimovic1997c, dorelliscudder1999, Lamy2003, Zouganelis2004a}. 

On the other hand, observations have shown that the relative density of the two high-energy electron populations exchanges as a function of radial distance. The strahl is more pronounced close to the Sun while the halo density increases over the radial distance \citep{Stverak2009a}. This suggests that the halo is not present in the solar corona and is formed during the solar wind expansion from the strahl component.

The strahl and the halo populations, not sensitive to collisions, were early assumed to be the remnant of the hot coronal electrons in the solar wind \citep{Feldman1975a}. The focusing mechanism experienced by the strahl during the expansion does not affect the shape of the magnetic field aligned cut through the strahl VDF (f$_{s\parallel}$) nor the strahl parallel temperature (T$_{s\parallel}$). Therefore, the strahl in absence of collisions any other interactions preserves the temperature and the shape of the VDF of the coronal electrons at its origin. 

This is only valid in the kinetic models not including collisions or wave particle interactions. The strahl electrons have been observed to not focus, but scatter with radial distance \citep{Hammond1996b, Graham2017a,Bercic2019} accounting this phenomena to some extent to Coulomb collisions \citep{Horaites2018a, Horaites2019}, but also to wave-particle interactions \citep{Vocks2005a, Kajdic2016a} and scattering by the background turbulence \citep{Pagel2007a, Saito2007a}. \citet{Graham2017a} report that the strahl was rarely observed at the distances higher than 5 au. The strahl and the halo electrons do interact with the surrounding plasma and electric and magnetic fields, but on much larger spatial scale than the thermal, core electron component. 

The core electron temperature was recently found to be correlated to the solar wind origin in the inner heliosphere, however, the correlation is almost completely lost by the time the solar wind reaches the distance of 1 au (Maksimovic et al, 2019).

Whether the high-energy electron components preserve information about the solar wind origin at the radial distance of 1 au has been tested through comparison to the oxygen charge state ratio (O$^{7+}$/O$^{6+}$), an established proxy for measuring the coronal electron temperature. While  \cite{Hefti2008} find a correlation between the T$_{s\parallel}$ and the oxygen charge state ratio, \citet{MacNeil2017} find that the correlation is not very strong and it varies depending on the choice of interval. 

We aim to investigate whether the information about the solar wind origin is present at the closest distances sampled by in-situ instruments so far:  35 R$_S$ for the Parker Solar Probe (PSP) and 65 R$_S$ for the Helios mission. As the oxygen charge state ratio is not measured by these two space crafts we use the solar wind velocity as an indicator of the solar wind origin.

\section{Data sets}
\subsection{Parker Solar Probe}
\label{sec:data-psp}
Launched in August 2018, PSP \citep{Fox2016a} is a mission designed to study the solar wind in the vicinity of the Sun, eventually reaching as close as 8.8 R$_S$ from its surface. We analyse the data gathered during the first two orbits of PSP with the perihelion of 34.7 R$_S$ and the aphelion between the orbits of Venus and Earth.

\begin{figure*}
\centering
\includegraphics[width=\hsize]{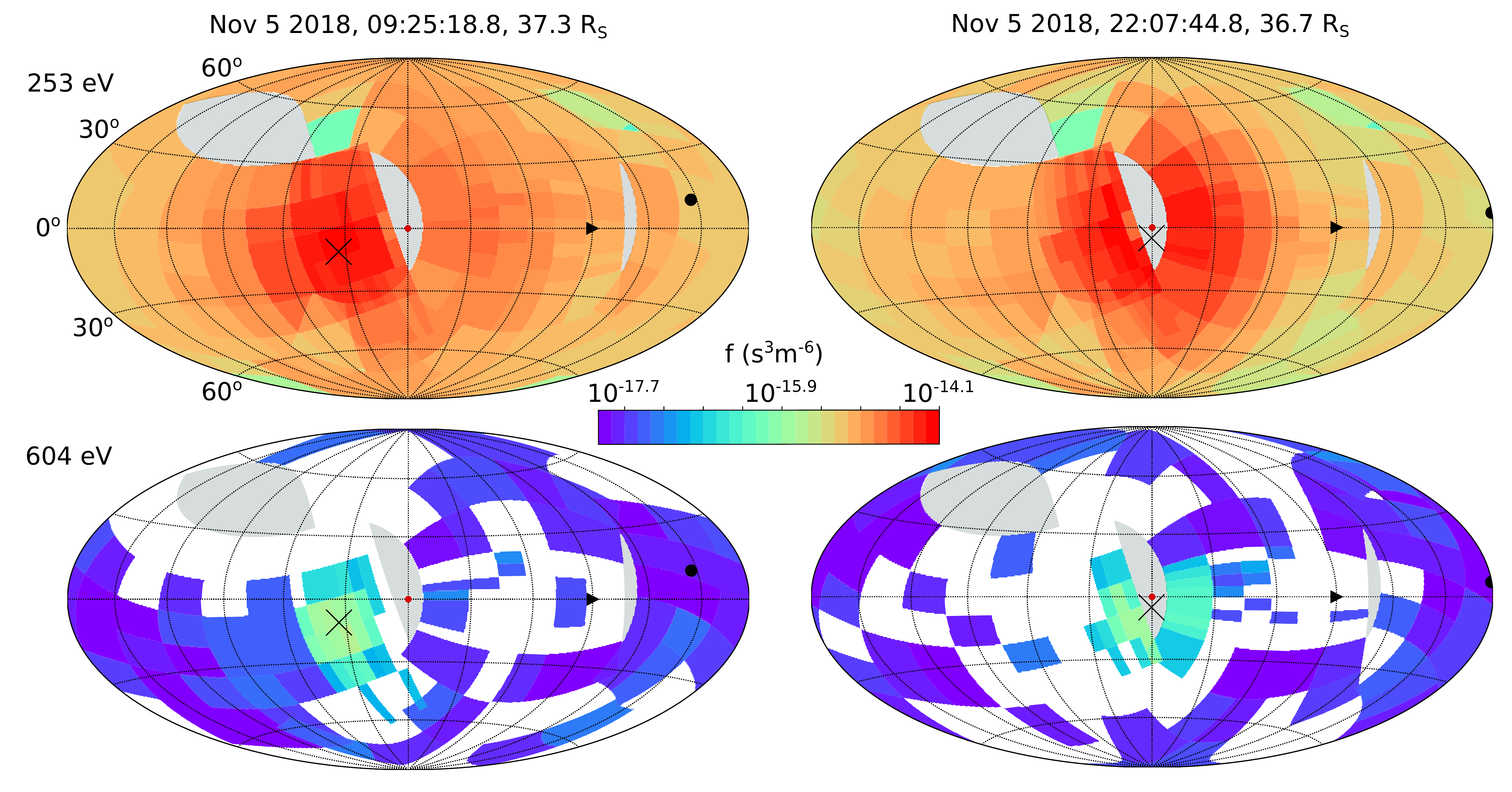}
\caption{Combined SPAN-E FOVs showing two examples (columns) of a full angular scan for two energy bins (rows). The examples (left - Nov 5 2018, 9:25:18, right - 22:07:44) were selected due to their different orientation of magnetic field in the FOV. A colour denotes the value of the VDF in each angular bin. The horizontal axis of FOVs is aligned with the spacecraft orbital plane. The Sun-spacecraft line is marked with the red dot and is in the middle of each plot. Vertical dimension thus shows angles out of orbital plane. The spacecraft is moving toward the black triangle, and the black dot and the black cross denote magnetic field positive and negative directions. The light grey areas represent the solid angles not sampled by the two instruments. 
}
\label{fig:fov}%
\end{figure*}

Electrons on-board PSP are measured with two SPAN Electron (SPAN-E) electrostatic analysers: SPAN-A and SPAN-B \citep{Whittlesey2020}, part of the SWEAP instrument suite \citep{Kasper2016}. Positioned on the ram and on the anti-ram side of the spacecraft with their 120$^o$ $\times$ 240$^o$ field of views (FOV) 90$^o$ tilted with respect to each other, they cover almost full 4$\pi$ solid angle. The azimuth angle ($\phi$) on each of the SPAN-Es is measured by 8 small (6$^o$) and 8 large (24$^o$) anodes, while the elevation ($\theta$) angles are sampled by the electrostatic deflectors. During the first two encounters deflectors separated the elevation measurements in 8 angular bins with a resolution of 20$^o$, of which the two extreme elevation bins have not been used in our analysis. The combined FOV of the two instruments is represented in Figure \ref{fig:fov}, where the grey surfaces represent solid angles which are not sampled by the instruments. To be able to withstand high levels of solar radiation, PSP is equipped with a heat shield. When the spacecraft is within 0.7 au from the Sun, the shield points straight to it and blocks approximately an angle of 10$^o$ from the Sun-spacecraft line (the centre of the FOVs in Fig. \ref{fig:fov}). Electron energy is measured by toroidal electrostatic analyzers, which are adapted to the high variation of electron fluxes with a mechanical attenuator controlling the size of the entrance to the aperture. Energies between 2 eV and 2 keV are sampled in 32 exponentially spaced bins with the energy resolution ($\Delta$E/E) of 0.07.

The duration of one sweep over all the energy and deflection bins is 0.218 s. The data product used for the presented data analysis are full 3D spectra (32 energies, 8 elevations, 16 azimuths) integrated over a period of 27.9 s during Encounter 1 (Oct 29 - Nov 14 2018) and over a period of 14.0 s during Encounter 2 (Mar 29 - Apr 10 2019). When the spacecraft is further from the Sun ($>$ 60 R$_S$) the instruments are operating in cruise mode with the cadence of 895 s and integration period of 27.9 s.

Detailed descriptions of the SPAN-E instruments and their operating modes are provided by \citet{Whittlesey2020}.

In addition to the electron measurements we use the solar wind proton velocity and density moments calculated from the SPC instrument \citep{Case2019} and a vector magnetic field measured by a triaxial fluxgate magnetometer MAG part of the FIELDS investigation \citep{Bale2016}. SPC is a Faraday cup instrument sticking out of the heat shield and measuring the plasma flowing directly from the Sun, also part of the SWEAP investigation \citep{Kasper2016}. The cadence of both, SPC and MAG, is higher than that of SPAN-E, thus the averages over the duration of each full SPAN spectra are used in further analysis.

\subsection{Helios 1}
The predecessors of the PSP are the two Helios missions launched in the 70s \citep{Porsche1981a}. For more than 6 years these two spacecraft were exploring the inner heliosphere down to 0.3 au (64 R$_S$) and provided us with a big data set of various solar wind parameters, among others revealing radial and solar cycle related trends\citep{Feldman1975a,Pilipp1987b,Maksimovic2005a,Marsch2006,Stverak2009a}. These data were of great importance during the preparation for the PSP mission and stay important due to the large statistics and radial and time coverage. In this work we use the data from Helios 1 gathered between 1974 and 1980.

Electron VDFs on-board Helios 1 mission are sampled by a single narrow 2$^o$ $\times$ 19$^o$ FOV aperture, which uses spacecraft spin to obtain a 2D measurement in the plane perpendicular to the spin axis. The sampled plane is aligned with the ecliptic plane. The 360$^o$ azimuth angle measurement is completed in 8 steps resulting in 28.1$^o$ wide azimuth bins with gaps in between them (see schematics in Fig. \ref{fig:helios} (a)). Energies between 9 eV and 1.5 keV are sampled in 16 exponentially spaced energy steps. The full 2D measurement (16 energies, 8 azimuths) is completed in 16 s with a cadence of 40 s.

The proton on-board integrated densities and velocity vectors were taken from the original Helios files in Helios data archive \footnote{Link to the data archive: \emph{http://helios-data.ssl.berkeley.edu}}. 

The magnetic field vector is a composite measurement of two fluxgate magnetometers: E2 for all instances where measured magnetic field was less than 50 nT, and E3 for the rest. More details about the Helios data set and instrumentation can be found in our previous work with Helios observations \citep{Bercic2019}.

\section{Method}
\label{sec:method}

\subsection{Parker Solar Probe}

The measured electron distribution functions are subject to instrumental as well as environmental effects. An important issue on the instrumental side is the determination of sensitivities of each of the azimuth anodes. The sensitivity coefficients used for our analysis were obtained through in-flight calibration described in the work of \citet{Halekas2019}. The effects of the spacecrafts own magnetic field and electric charge on the particle trajectories were studied by \citet{McGinnis2019}. They show that, even though the spacecraft magnetic field is relatively large (it was predicted to reach the strength of 500 nT), the effect on some of the plasma moments, is small (see Table 2 in \citet{McGinnis2019}). The biggest errors were found for the bulk velocity calculation as it strongly depends on low energy measurements. The smallest errors, on the other hand, arise for the temperature calculation more dependent on higher energy measurements. The spacecraft potential was estimated to be low, on the order of a few Volts negative during the first two encounters. As our main focus in this article are the high energy (strahl) electrons, we believe that our results are not affected significantly by these effects which are more relevant for the low energy electrons \citep{Salem2001}.

The instruments' lower energy bins are contaminated by secondary electrons emitted from the spacecraft. \citet{Halekas2019} choose to include them in their fitting model as a Maxwellian distribution with a temperature of 3.5 eV. For the purpose of our work we find that it is sufficient to simply neglect the contaminated lower energy measurements. \\

We start our analysis with a rotation of the SPAN-A and -B velocity vectors from their initial instrument frame to the common RTN (Radial-Tangential-Normal) coordinate frame. In this frame R-axis is aligned with the Sun-spacecraft line and pointing away from the Sun, T-axis perpendicular to R-axis and pointing in spacecraft ram direction and N-axis completing the right-handed frame. The spacecraft velocity and the solar wind proton velocity as measured by SPC are then subtracted to shift the VDFs in the plasma rest frame. After that, the magnetic field measurement averaged to the SPAN full scan duration is used to rotate the VDFs to the magnetic field aligned frame. 

\begin{figure}
\centering
\includegraphics[width=\hsize]{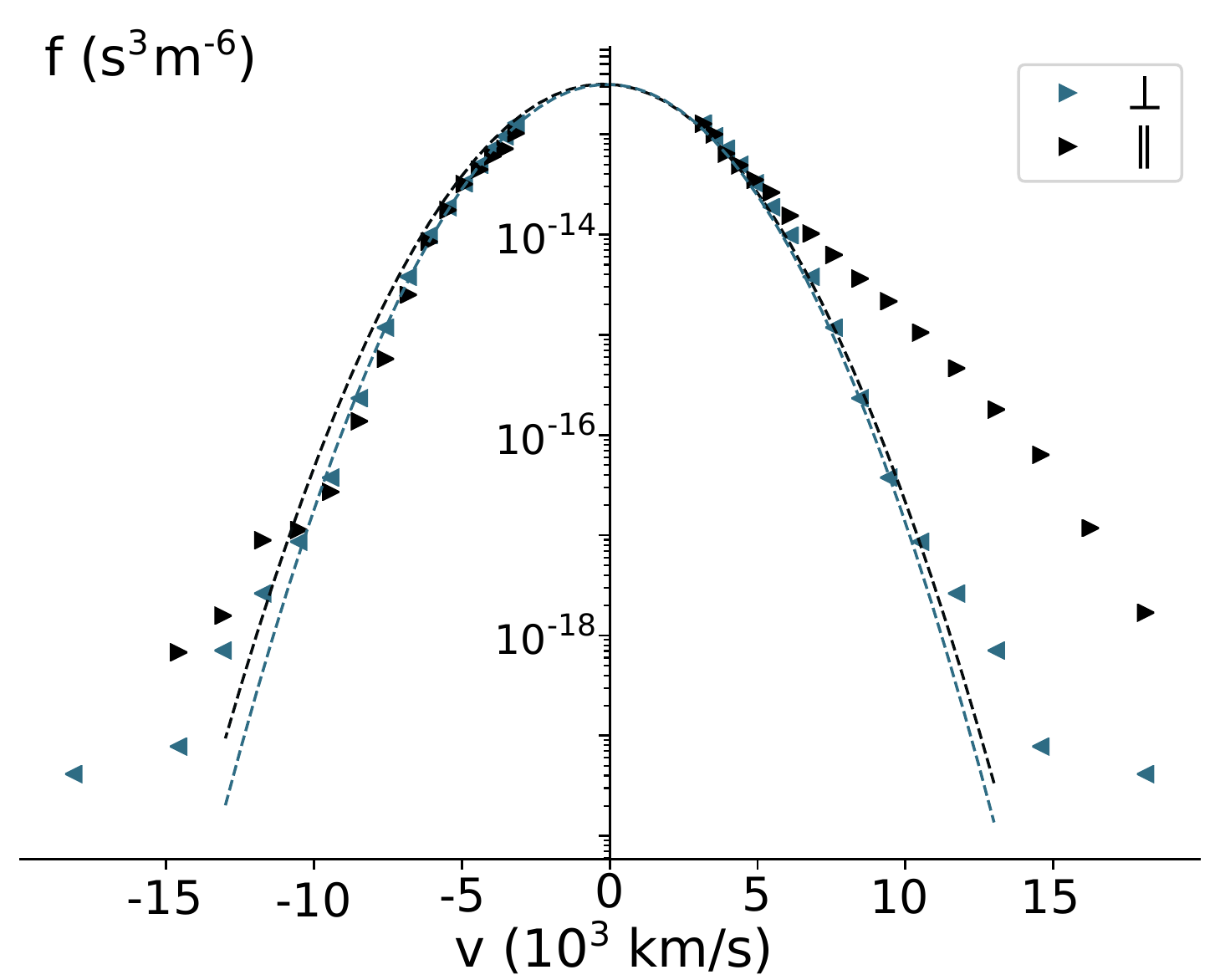}
\caption{Parallel ($\parallel$) and perpendicular ($\perp$) cuts through an electron VDF measured by SPAN-E instruments on Nov 5th at 9:25:18 (the same example as on the left side of Fig. \ref{fig:fov}). The positive velocity values for the parallel cut represent the part of the distribution aligned with the magnetic field and directed in the anti-sunward direction. The perpendicular values are the same on both sides of the plot as there is no preferred direction perpendicular to magnetic field. The data points presented with rightward pointing triangles ($>$) were provided by SPAN-A, while the leftward pointing triangles ($<$) represent the points from SPAN-B instrument. The strahl electrons in this scan are detected by SPAN-A agreeing with the FOV representation in Fig. \ref{fig:fov}.  }
\label{fig:vdfcut}%
\end{figure}

Following the works of \citet{Maksimovic1997b, Stverak2009a, Bercic2019} the core electrons are modelled with a 3-dimensional bi-Maxwellian distribution function:
\begin{multline}
f_c (v_{\perp1},v_{\perp2}, v_\parallel) =  \\
= A_c \exp \Big( \frac{(v_{\perp1}-\Delta v_{\perp1})^2}{w_\perp^2} + \\
+ \frac{(v_{\perp2}-\Delta v_{\perp2})^2}{w_\perp^2} + \frac{(v_\parallel-\Delta v_\parallel)^2}{w_\parallel^2} \Big)
\end{multline}

where $\Delta v_{\perp1, \perp2, \parallel}$ are the drift velocities corresponding to three axes of the magnetic field aligned frame. The fits were preformed on the full 3-dimensional VDFs using a least-square minimisation algorithm\footnote{scipy.optimize.leastsq (https://docs.scipy.org/doc/scipy/
reference/generated/scipy.optimize.leastsq.html)} provided by Scipy Optimization package for Python programming language \citep{virtanen}. Because the VDF values span over several orders of magnitude (see Fig. \ref{fig:vdfcut}) the fitting was carried out in logarthimic space ($\ln(f_c)$). This technique decreases the large difference in the weight of fitted data points, giving more importance to the low VDF values. From our 6 fitting parameters - $A_c$, $w_\perp$, $w_\parallel$, and $\Delta v_{\perp1, \perp2, \parallel}$ - we can obtain the core density $n_c$ from:
\begin{equation}
n_c = A_c \cdot \pi^{3/2} w_\perp^2 w_\parallel.
\end{equation}

The thermal speeds parallel ($w_\parallel$) and perpendicular ($w_\perp$) to the magnetic field can be expressed in terms of core temperature $T_{c\perp,\parallel}$:
\begin{equation}
    T_{c \perp,\parallel} = \frac{m_e w_{\perp,\parallel}^2}{2k_B},
\end{equation}
where $k_B$ is Boltzman constant and $m_e$ mass of an electron. The core density and parallel temperature are then used to calculate the electron parallel plasma beta parameter:
\begin{equation}
\beta_{ec\parallel} = \frac{2 \mu_0 n_c k_B T_{c \parallel}}{B^2}, 
\label{eq:ecpar}
\end{equation}
with $\mu_0$ standing for vacuum permeability and $B$ for magnetic field.

An example of electron VDF measured on the Nov 5th is presented with the cuts through the parallel ($\parallel$) and the perpendicular ($\perp$) direction with respect to the magnetic field in Fig. \ref{fig:vdfcut}. We recognise the expected electron VDF features: a core fitted with a bi-Maxwellian distribution (dashed line in the Fig. \ref{fig:vdfcut}), a field aligned strahl component only seen parallel to the magnetic field direction, and a weak halo departing from a Maxwellian fit at higher electron energies. Another feature we do not plan to discuss in the present work, already observed by \citet{Halekas2019}, can be recognised in Fig. \ref{fig:vdfcut}. Directed towards the Sun (on the left side) and aligned with the magnetic field (dark blue) there appears to be a deficit in the core electron distribution; a part of phase space where the measured VDF appears to be smaller than the best fitting Maxwellian distribution function. \\

Even though the two SPAN-E instruments cover almost a full solid angle, there exist cases when the electron VDFs are not fully characterised by the measurement. As introduced in Sec. \ref{sec:intro}, we investigate the behaviour of the strahl electrons, a population aligned with the magnetic field and directed away from the Sun. The magnetic field closer to the Sun fluctuates around a vector more and more aligned with the radial direction following the Parker spiral model \citep{Parker1958}. This means that often the magnetic field measurement over one full SPAN-E scan duration will lie in the portion of the FOV where the solar wind electrons are blocked by the spacecraft heat shield (marked with grey in the centre of the FOVs in Fig. \ref{fig:fov}). A case when this happens is shown on the right side of Fig. \ref{fig:fov}. At lower energies where the width of the strahl electron beam is larger (upper FOV: 253 eV) the effect of the FOV obstruction does not play a big role, while at high electron energies (lower FOV: 604 eV) where the strahl electron population often appears very narrow we might be missing a big part of the strahl VDF. An opposite case, when the strahl is detected as accurate as possible is presented on the left side of Fig \ref{fig:fov}. When the magnetic field direction lies within the area of the FOV covered by the small anodes of the SPAN-A the strahl electrons are measured with the angular resolution of 6 $\times$ 20$^o$ (azimuth $\times$ elevation)\citep{Whittlesey2020}. We do not wish to limit our data set with respect to the magnetic field direction because we expect that the physical mechanisms shaping the electron VDFs will also depend on magnetic field vector. Instead we use a fitting method described below which accounts for the field of view limitation. The differences resulting from the FOV obstruction are further analysed and presented in Appendix \ref{app:fov}.

\begin{figure}
\centering
\includegraphics[width=\hsize]{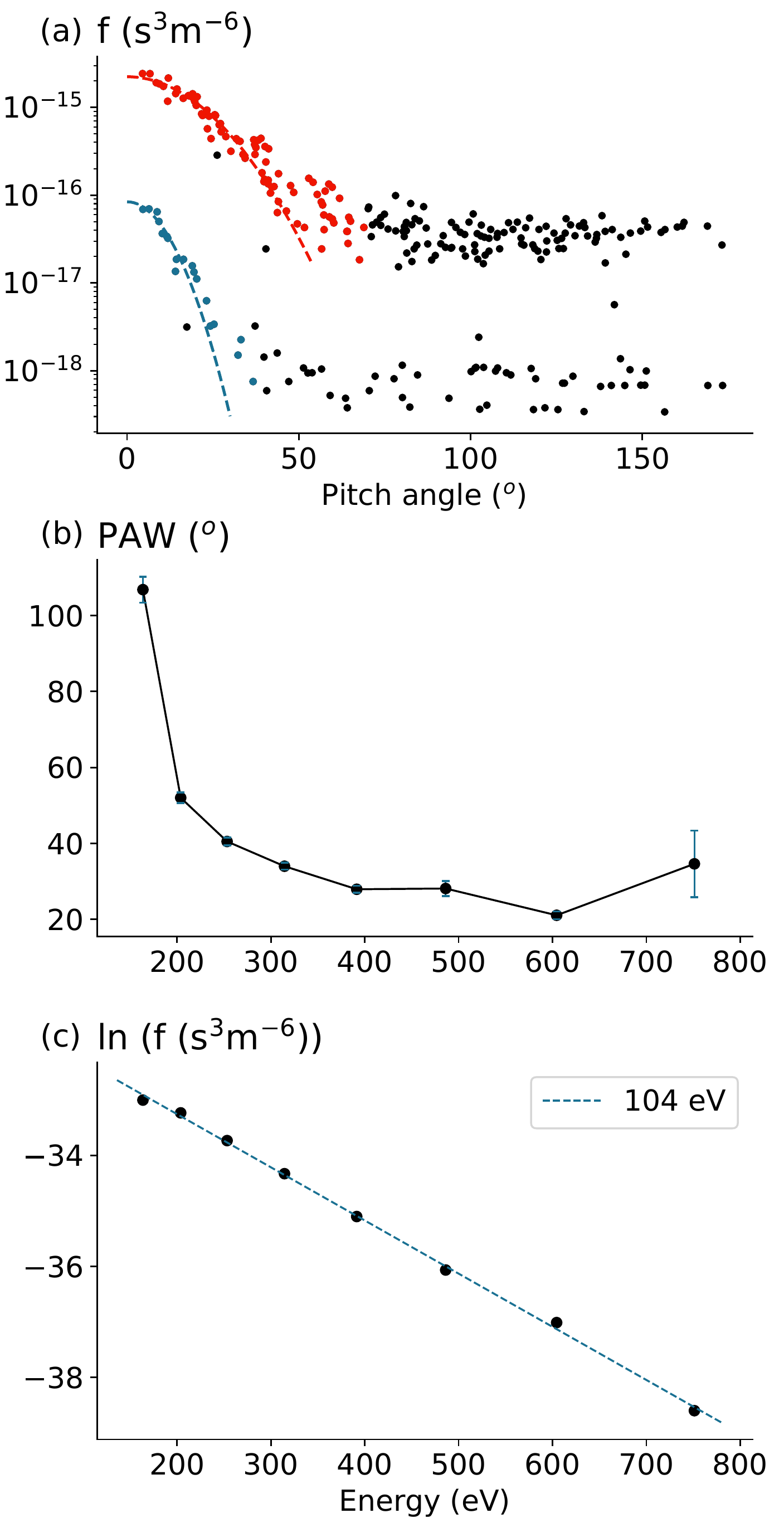}
\caption{An example illustrating the strahl characterisation method. All three plots come from one SPAN-E full spectra measurement, the same as shown in Fig. \ref{fig:vdfcut} and left of \ref{fig:fov}. 
(a) Pitch angle distributions shown for two different energy bins (253 eV in red and 604 eV in blue) with fitted normal functions (Eq. \ref{eq:pa}) marked with dashed lines. The points used calculation of PAW and f$_{max}$ are marked with red and blue, and the background in black. The obtained PAWs for these two energy bins were 40$^o$ and 22$^o$.
(b) Strahl PAW (Eq. \ref{eq:paw}) calculated for each of the energy bins. The error bars denote an interval of one standard deviation.
(c) Natural logarithm of the f$_{max, i}$ plotted against the electron energy and the linear fit preformed in this parameter space (dashed line) to obtain the strahl parallel temperature (T$_{s\parallel}$) in this example resulting to 104 eV (see Eq. \ref{eq:lnfit}).}
\label{fig:pawfit}%
\end{figure}

We characterise the strahl electrons with two parameters: strahl pitch-angle width (PAW) and strahl parallel temperature (T$_{s\parallel}$). 

We expect to observe the strahl component aligned with the magnetic field and moving away from the Sun. This means that if the magnetic field radial component is negative the strahl electrons will be anti-parallel to the magnetic field vector. However, this is not always the case. Bi-directional strahls have been observed and related to magnetic field structures like closed magnetic loops and magnetic clouds \citep{Gosling1987}. Sunward directed strahls have also been observed and serve as the indicators of magnetic field structures sometimes referred to as the switchbacks \citep{Balogh1999,Yamauchi2004,MacNeil2017}, which are frequently observed also during the first perihelion of the PSP \citep{Kasper2019, Bale2019}. In this study we do not consider special cases and focus on the anti-sunward moving strahl electrons in the nominal solar wind.

To obtain the strahl PAW we first calculate pitch angles ($\alpha$) for each measured energy bin of the electron VDF put in the plasma rest (defined by the solar wind protons) and magnetic field aligned frame using the following criteria:
\begin{eqnarray}
\text{if }B_r > 0:\nonumber \\  
    \alpha (v_{\perp1},v_{\perp2}, v_\parallel) = \arccos { \big( \frac{v_\parallel}{\sqrt{v_{\perp1}^2 + v_{\perp2}^2}} \big) },\\
\text{if }B_r < 0:  \nonumber \\ 
    \alpha (v_{\perp1},v_{\perp2}, v_\parallel) = \arccos { \big( \frac{-v_\parallel}{\sqrt{v_{\perp1}^2 + v_{\perp2}^2}} \big) }.
\end{eqnarray}

The obtained pitch angles ($\alpha_i$) thus lay on the interval [0$^o$, 180$^o$], where 0$^o$ denotes the direction along the magnetic field and pointing away from the Sun, 90$^o$ direction perpendicular to the magnetic field, and 180$^o$ the direction along the magnetic field and pointing towards the Sun.

These pitch-angle distributions are then fitted for each energy bin separately with a Gaussian distribution function:
\begin{equation}
f_i(\alpha) = f_{max,i} \cdot \exp \big(\frac{\alpha^2}{2 \sigma_i^2}  \big),
\label{eq:pa}
\end{equation}
where subscript $i$ denotes iteration over all the energy bins. 
Two fits are preformed for each energy bin. A first fit to all the available points in an energy bin to separate the strahl from the background, and a second fit only to the points aligning with the first fit selected by the following criteria:
\begin{equation}
|(f_{data} - f_{fit,1})| / f_{data} < 0.99,
\label{eq:criteria}
\end{equation}
where $f_{data}$ are the data points and $f_{fit,1}$ the values predicted by the first fit. The second fit was performed when at least 4 data points conform to the criteria above (Eq. \ref{eq:criteria}).
Two examples of the second fit are shown in Fig. \ref{fig:pawfit} (a) (dashed lines). The selected points representing the strahl part of the distribution are marked with red or blue colour. We use the parameters from the second fit to then calculate the full-width-half-maximum (FWHM) which we refer to as the pitch-angle width (PAW):
\begin{equation}
\text{PAW}_i = 2 \sqrt{2 \ln{2}} \cdot \sigma_i.
\label{eq:paw}
\end{equation}
The instances when PAW exceeds the value of 180$^o$ are excluded from further analysis, as they indicate almost isotropic pitch-angle distributions and could be dominated by one of the more isotropic solar wind electron components, the core or the halo. In this work we choose to perform the PAW analysis on the full electron VDF and not only on the strahl VDF, which can be obtained by subtraction of the modelled core and halo components from the total measured VDF (as done by \citet{Bercic2019}, Sec. 3.2). Using the full VDF we avoid the errors resulting from the core and the halo modelling. The core population taking up the lower electron energy is more sensitive to the effects of spacecraft's magnetic and electric fields and exhibits the yet unstudied deficit in the sunward, magnetic field aligned portion of the VDF. The halo component is difficult to model because it was observed to be very tenuous and, during the encounter periods when the instrumental mechanic attenuator was closed, represented only by a few data points \citep{Halekas2019}. The strahl component, on the other hand, takes up higher energies and appears relatively dense, especially during the encounter periods. For these reasons we fit Eq. \ref{eq:pa} the full measured VDF and rely on the assumption that the energy bins resulting in PAW $<$ 180$^o$ are dominated by the strahl electron component. The same approach was used by \citet{Hammond1996b, Graham2017a}. An example of the PAWs calculated for each energy bin of one measured scan is shown in Fig. \ref{fig:pawfit} (b). \\

In the inner heliosphere and for the energy ranges sampled by the SPAN-E instruments the strahl VDFs along the parallel direction to the magnetic field are well represented by a Maxwellian distribution function. For the scope of this work we are only interested in the temperature of this Maxwellian - the slope of the parallel strahl VDF. However, the peak of the pitch-angle distributions aligned with magnetic field is sometimes not sampled due to the heat shield FOV obstruction. Thus instead of using measured VDF closest to the parallel direction, we use the maximum VDF values ($f_{max,i}$) from the fit to the pitch angle distributions at each energy bin (see Eq. \ref{eq:pa}).

We perform a fit in the parameter space where a Maxwellian distribution forms a straight line with a slope depending only on its temperature:

\begin{equation}
\ln {f_{max,i} (v_\parallel)} = -\frac{m_e}{2k_B \cdot T_{s\parallel}}\cdot v_\parallel^2 + ln \big( n_s \sqrt{\frac{m_e}{2 \pi k_B \cdot T_{s\parallel}}} \big),
\label{eq:lnfit}
\end{equation}

where $ln$ denotes the natural logarithm. An example of the strahl distribution in this representation is shown in Fig. \ref{fig:pawfit} (c). Fitting a straight line in this parameter space we assume that the drift velocity of the Maxwellian is 0, or very small in comparison to electron velocity ($v$). This agrees with the exospheric models predicting the VDF to stay the same as in the corona, where the bulk velocity of electrons is 0.

\subsection{Helios 1}
The same two parameters to characterise the strahl electrons, PAW and $T_{s\parallel}$, were obtained from the Helios 1 observations. The PAWs as well as some other parameters like the core electron density ($n_c$) and temperature ($T_{c\parallel, \perp}$), velocity of the protons ($v_p$), and magnetic field (B), have already been used and are described in our previous work \citep{Bercic2019}. 

\begin{figure*}
\centering
\includegraphics[width=\hsize]{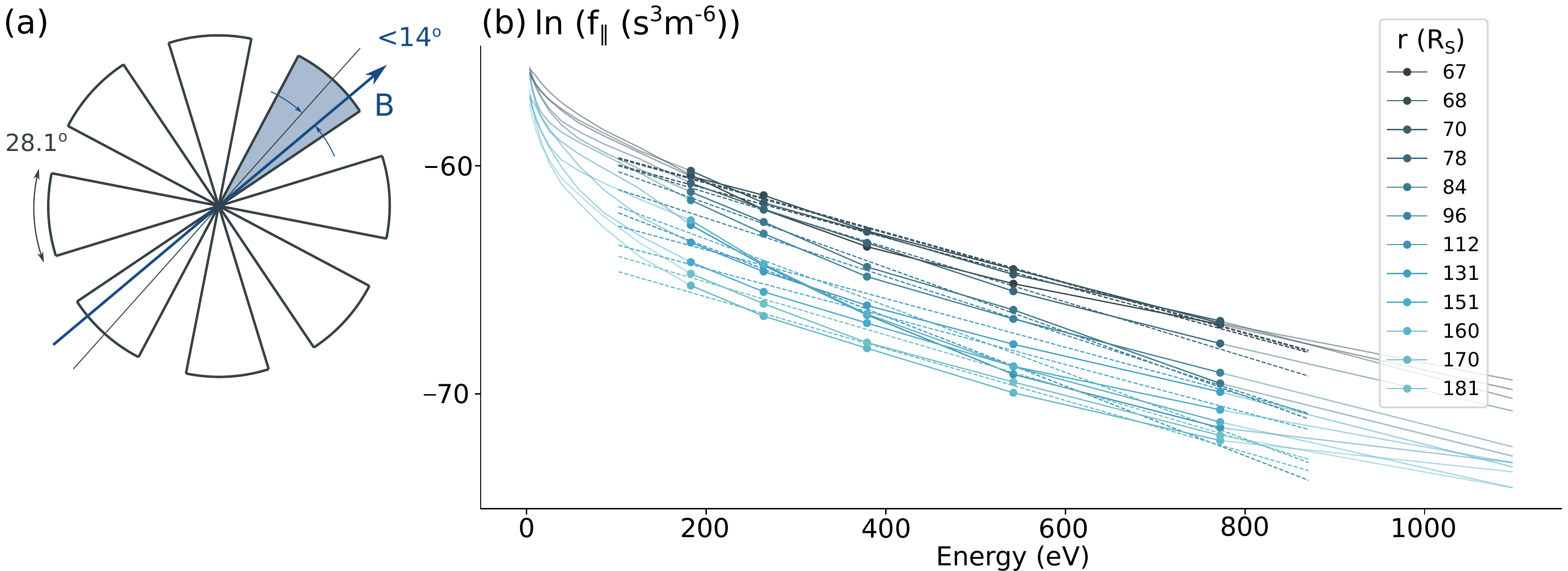}
\caption{(a) Schematics of the electron instrument on-board Helios 1 mission. The instrument has 8 azimuth bins which are 28.1$^o$ wide and separated with gaps. With B we mark the magnetic field direction. (b) Each line represents an electron VDF detected by the azimuth bin aligned with the magnetic field direction (marked with blue in the schematics) and averaged over 10 consecutive measurements. We compare VDF examples from a half of Helios 1 orbit between Sep 21 and Dec 8 1975 spanning distances from 67 to 181 R$_S$ (see the legend). The dots mark the measurements used for the $T_{s\parallel}$ fits, which are shown with the dotted lines.}
\label{fig:helios}%
\end{figure*}

The strahl temperature is determined in a similar way as described for the PSP, however, the Helios 1 mission did not have a heat shield and the 2D electron instrument was able to point straight at the Sun, thus using the $f_{max,i}$ parameter from the PAW fits is not necessary. Instead we limit the data set to instances when magnetic field lies within one of the eight azimuth bins and fit the Eq. \ref{eq:lnfit} to the data points from this azimuth bin (marked in blue in Fig. \ref{fig:helios} (a)). We use a full measured distribution function in this bin, and not the strahl distribution presented in \citet{Bercic2019}, which was obtained by subtraction of the core and the halo from the measured VDF ($f_{strahl} = f_{measured} - f_{core} - f_{halo}$). The full distribution was used to unify the methods of $T_{s\parallel}$ calculation between PSP and Helios data set.

This process is illustrated in Fig. \ref{fig:helios} (b). From the Helios data set, it is not as obvious that the strahl parallel VDF can be modelled by a Maxwellian. The VDFs appear noisier and especially further from the Sun (lighter blue values Fig. \ref{fig:helios} (b)) may exhibit traces of high energy tails, previously modelled by Kappa-like distribution functions \citep{Maksimovic1997c, Stverak2009a}. However, for a certain energy range (between $\sim$ 200 and 800 eV) strahl VDFs still present a straight line in the $\ln {f} (v^2)$ parameter space and give us the information about the strahl parallel temperatures. Fitting only the selected energy range we avoid the inclusion of the electron core component.

\section{Observations}
\label{sec:observations}

\begin{figure*}
\centering
\includegraphics[width=1\hsize]{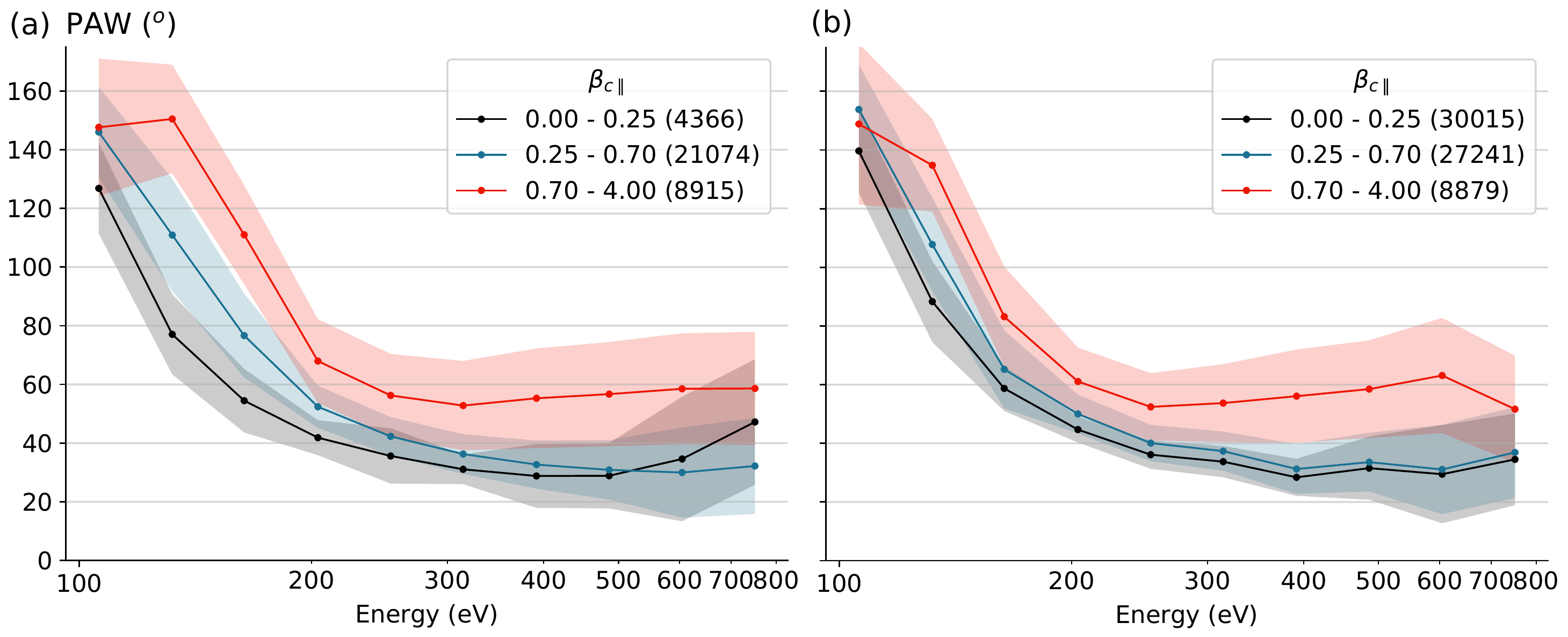}
\caption{Strahl PAW with respect to electron energy, separated into three bins according to the local $\beta_{ec\parallel}$ value marked in the legend. The number in parenthesis denotes a number of VDFs in each $\beta_{ec\parallel}$ bin. The shaded region for each line gives the span of one standard deviation. (a) Encounter 1, (b) Encounter 2.}
\label{fig:paw-beta}%
\end{figure*}

Strahl PAWs with respect to electron energy for the different plasma $\beta_{ec\parallel}$ values are shown in Fig. \ref{fig:paw-beta}. Only data gathered during the first two encounter periods (35 - 60 R$_S$) was used and plotted separately ((a) - encounter 1, (b) - encounter 2). We separated the data because of the different integration time of the instruments for each encounter (see Sec. \ref{sec:data-psp}), and because of an unresolved issue with the instruments response during the encounter 2. This artefact can be seen in Figs. \ref{fig:paw-beta}(b),  \ref{fig:paw-fov}(b), and \ref{fig:paw-magstd}(b) as a zig-zag pattern of PAW along the energy dimension for higher energies. It appears as if the PAW is slightly broader for every second energy bin. A possible reason for this kind of measurement response could lie in the hysteresis of the instrument deflection plates. A predicted correction for this effect has been applied on the whole data set, however, the hysteresis could be time dependent with a stronger effect on the measurements made during the encounter 2. Nevertheless, the irregularities do not exceed the statistical error and thus do not change any conclusions of the present work.

Both plots in Fig. \ref{fig:paw-beta} show the increase of PAW with $\beta_{ec\parallel}$. For the lower two $\beta_{ec\parallel}$ cases, the PAW decreases with electron energy reaching down to 30$^o$, while for the highest $\beta_{ec\parallel}$ case the PAW stays more or less constant with a value $\sim$ 55$^o$ for the energies above $\sim$ 200 eV. This high-$\beta_{ec\parallel}$ regime where the strahl appears to be more affected by the scattering mechanisms was found for 26 \% of electron spectra during the encounter 1 and for 13 \% during encounter 2. For all cases a fast increase of PAW is observed for the low electron energies, denoting the presence of the electron core population below the energy of $\sim$ 200 eV .

No radial dependence was found during the encounter periods (from 35 to 60 R$_S$), most likely as a consequence of the for now limited PSP data set. It appears that the type of the solar wind we observe has more effect on PAW than the radial distance. \\

\begin{figure*}
\centering
\includegraphics[width=1\hsize]{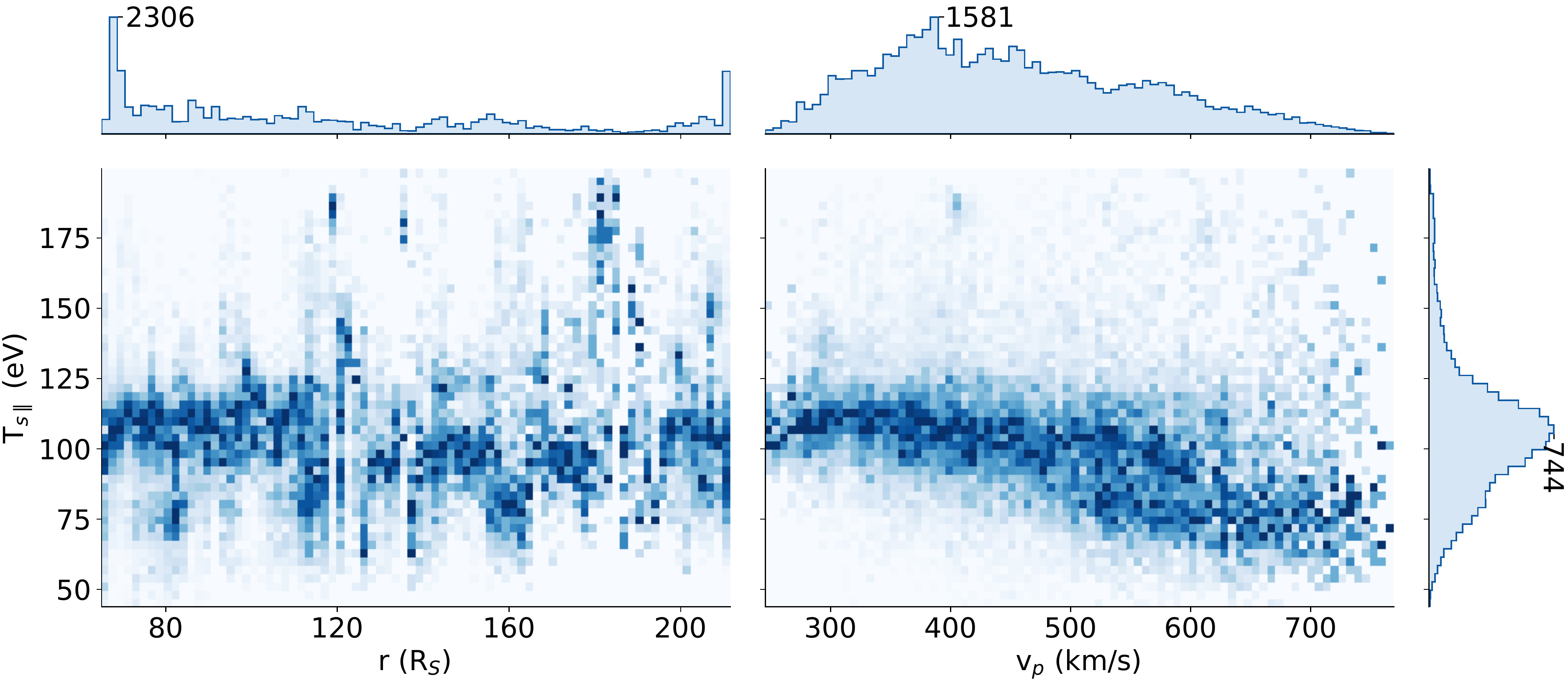}
\caption{Histograms showing the variation of T$_{s\parallel}$ with radial distance (r) (left), and solar wind proton velocity (v$_p$) (right) for the \emph{Helios} data set. The histograms are normalised to the maximum value in each vertical column. Above each of the 2D histograms and on the right side 1D histograms present the probability distribution of the corresponding parameters (r, v$_p$, and T$_{s\parallel}$). }
\label{fig:tsm-rad-helios}%
\end{figure*}

\begin{figure*}
\centering
\includegraphics[width=1\hsize]{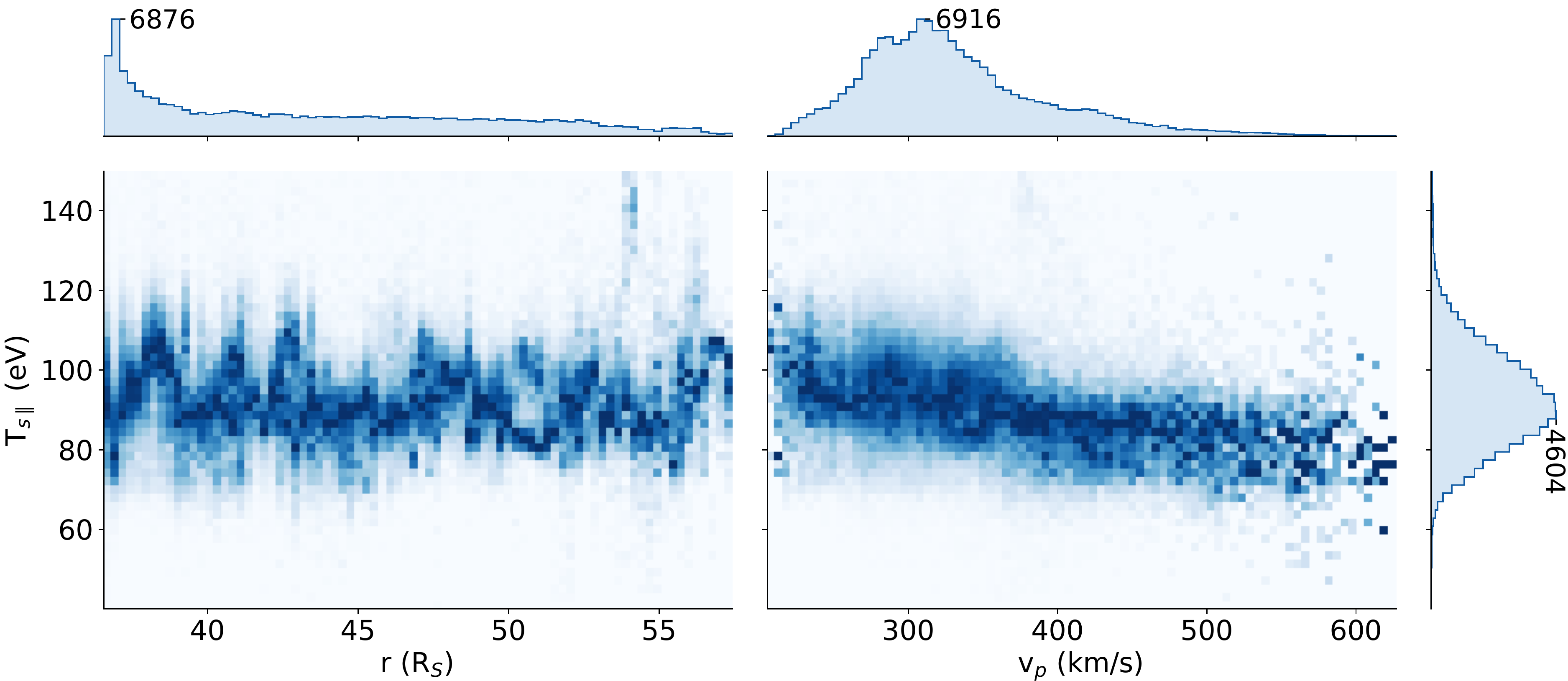}
\caption{Histograms showing the variation of T$_{s\parallel}$ with radial distance (r) (left), and solar wind proton velocity (v$_p$) (right) for the \emph{PSP} data set. The histograms are normalised to the maximum value in each vertical column.}
\label{fig:tsm-rad-psp}%
\end{figure*}

\begin{figure*}
\centering
\includegraphics[width=1\hsize]{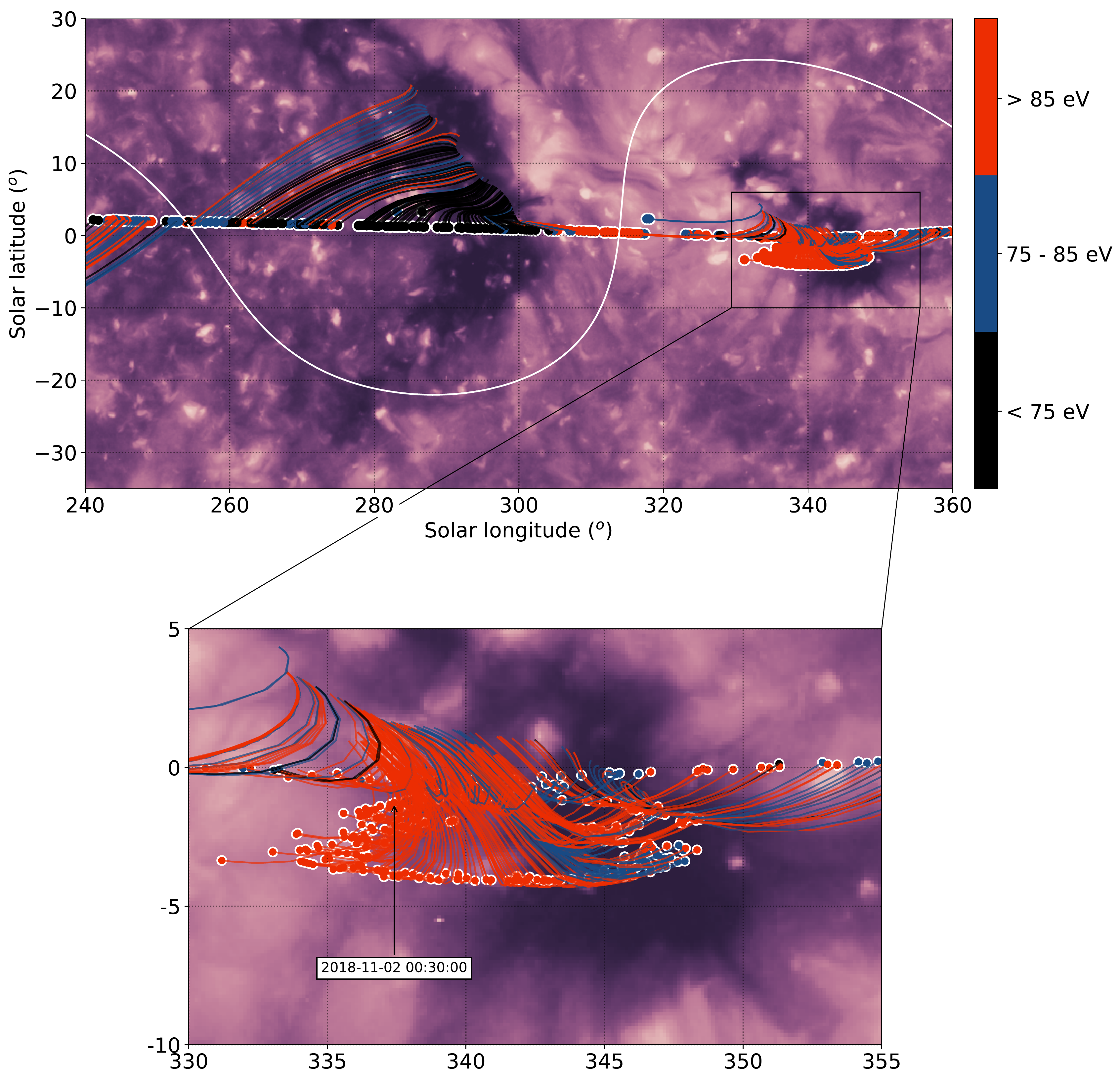}
\caption{The evolution of T$_{s\parallel}$ with part of the PSP orbit 1. The PSP trajectory is ballistically projected down to the corona (2 R$_S$) to produce sub spacecraft points. The coloured lines denote magnetic field lines mapped from the sub spacecraft points to the solar surface as predicted by the PFSS model with source surface height 2 R$_S$, the same as used in \citet{Bale2019, Badman2019}. The white line shows the PFSS neutral line. The points and magnetic field lines are coloured with respect to an hour long averages of T$_{s\parallel}$ (see the colour bar in (a)). The corresponding image of the Sun is a synoptic map of the 193 {\AA} emission synthesised from STEREO/EUVI and SDO/AIA for Carrington Rotation 2210, identical to the one used by \citet{Badman2019} in Figs. 5 \& 9. Upper plot presents a larger time interval (Oct 30 2018, 00:30 - Nov 23 2018, 17:30), and the lower a zoom of the encounter period (Oct 30 2018, 15:30 - Nov 14 2018, 8:30).}
\label{fig:tsm-sub-psp}%

\end{figure*}

Strahl temperatures obtained from Helios and PSP data sets are presented separately in Fig. \ref{fig:tsm-rad-helios} and \ref{fig:tsm-rad-psp}. The results from Helios data set are the outcome of binning several years of solar wind measurements while for the PSP we use the data obtained over less than 6 months. Nonetheless, during the PSP encounter periods the data rate is very high and we were lucky to have already sampled different types of solar wind providing us with a satisfactory statistics. For Helios data set sampled distances range from 65 to 215 R$_S$, while for the PSP the radial coverage is much smaller, from 35 and 58 R$_S$ (first two encounters). Similarly, measured proton velocities in the PSP data set have a smaller span than in the Helios data set. The 2D histograms in both cases show the same result, no strong trends in variation of the T$_{s\parallel}$ with radial distance (r) and an anti-correlation with the solar wind proton velocity (v$_p$). The overall mean value of T$_{s\parallel}$ measured by PSP is 93 eV with a standard deviation of 13 eV, and by Helios 105 eV with a standard deviation of 23 eV.

Fig. \ref{fig:tsm-sub-psp} presents the evolution of T$_{s\parallel}$ with part of the PSP orbit 1 trajectory ballistically projected down to the corona (2 R$_S$) to produce sub spacecraft points (marked with coloured dots). SWEAP in situ proton velocity measurements are used to perform this projection. The coloured lines show the magnetic field lines mapped from each of the sub spacecraft points down to the solar surface as predicted by the PFSS model (see \citet{Bale2019,Badman2019} for more details about the PFSS modelling). The polarity inversion line is shown in white.

This interval was chosen because it exhibits distinctive coronal features: a small coronal hole occurring during the first encounter period (region marked with a box in upper plot in Fig. \ref{fig:tsm-sub-psp}, enlarged in the lower plot), and a big coronal hole occurring after the encounter period (the centre of the upper plot in Fig. \ref{fig:tsm-sub-psp}). Coronal holes appear as darker parts in the images produced from the 193 {\AA} emission line, as these are the regions marked by low plasma density and open magnetic field lines. Oppositely, the bright regions in the image correspond to higher plasma densities, normally related to closed magnetic field loops. Similar plot has been shown in the work of \citet{Badman2019}, who use a PFSS model to map the magnetic field lines measured by the spacecraft back to the solar surface (see Figs. 5 \& 8 in the referred article). 

Our crude separation of T$_{s\parallel}$ appears to discern distinct coronal features as identified in the PFSS model: Very low T$_{s\parallel}$ (marked in black in Fig. \ref{fig:tsm-sub-psp}) is measured as PSP traces over the larger, positive polarity coronal hole after the first encounter and measured a fast wind stream, while a mix of intermediate T$_{s\parallel}$ (blue) and high T$_{s\parallel}$ (red) occurs in association with the smaller coronal hole PSP looped over at perihelion. The high strahl temperatures are associated with mapping to the edges of the coronal hole and proximity to the current sheet (white contour in Fig. \ref{fig:tsm-sub-psp}), while the intermediate temperature occurs at a time when the solar wind bulk speed increased and PSP was directly over the centre of the coronal hole. 

In lower plot in Fig. \ref{fig:tsm-sub-psp} presenting a zoom-in of the first encounter the intermediate T$_{s\parallel}$ do not correspond directly to the darker regions on the image. Coronal holes are dynamic features, and this small coronal hole has been observed to drift over the limb of the Sun on Nov 2 (a date marked in Fig. \ref{fig:tsm-sub-psp} (b)). Therefore, we can not be sure of position of the small coronal hole at the time of the PSP crossing and a slight disagreement between the image and T$_{s\parallel}$ is expected.

\section{Discussion}
\label{sec:discusion}

PAWs observed by the PSP agree very well with the reported observations from the Helios mission (Fig. 4 in \citet{Bercic2019}). The most obvious change in the radial evolution from the closest regions to the Sun probed by Helios spacecraft (65 R$_S$) to the first two perihelia of the PSP reaching down to 35 R$_S$ is the fast increase in PAW at low electron energies (see Fig. \ref{fig:paw-beta}). We attribute this increase to the presence of the core electron component reaching the temperatures above 30 eV \citep{Halekas2019}.

\begin{figure}
\centering
\includegraphics[width=1\hsize]{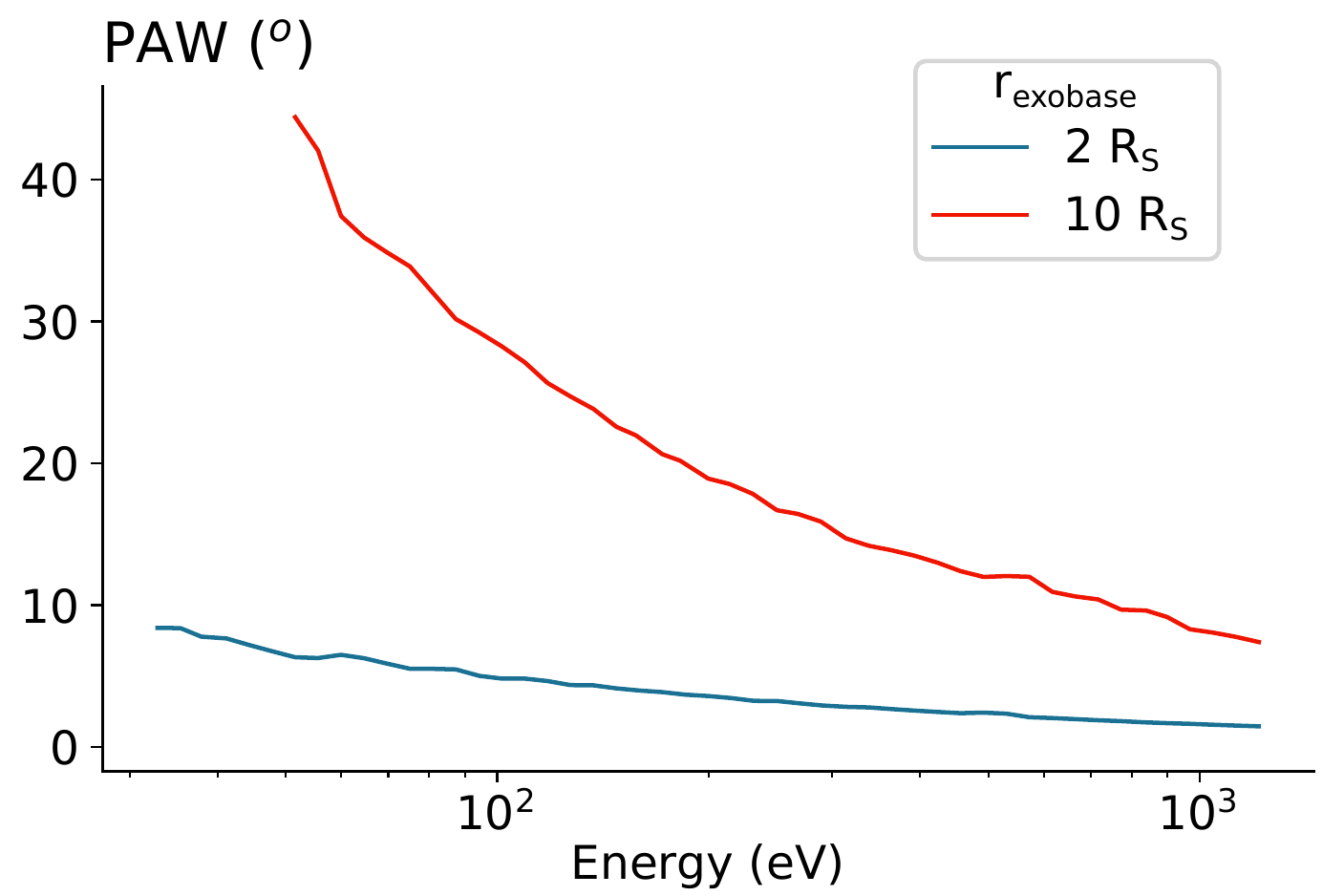}
\caption{Relation between PAW and electron energy at the distance of 35 R$_S$ resulting from a simple focusing model. The difference between the two curves is the selection of the exobase marked in the legend. The electric potential used for both examples was the same, equal to -500 V. }
\label{fig:focusing}%
\end{figure}

The observed anti-correlation between PAW and electron energy for the lower two $\beta_{ec\parallel}$ bins (Fig. \ref{fig:paw-beta}) might be a consequence of a collisionless focusing mechanism. Focusing of the solar wind electrons starts taking place at a distance above the solar surface where collisions can not dominate the electron VDF anymore, a distance in the frame of exospheric solar wind models referred to as the exobase \citep{Jockers1970, LemaireJosephandScherer1971}. The location of the exobase can be between 2 and 10 R$_S$ \citep{Maksimovic1997c}, and depends on the type of the solar wind. The expected relation between PAW and electron energy at the distance of 35 R$_S$ accounting only for the focusing mechanism is shown in Fig. \ref{fig:focusing} for the low and the high exobase limit. The model assumes an isotropic Maxwellian VDF at the exobase expanding along a radially decreasing magnetic field (B $\propto$ 1/r$^2$). Following energy and momentum conservation (Eq. 6 in \citet{Bercic2019}) we obtain a VDF at 35 R$_S$, and calculate the PAW as described in Sec. \ref{sec:method}. In comparison to the majority of observations the modelled PAWs still appear at least two times narrower.

\begin{figure}
\centering
\includegraphics[width=1\hsize]{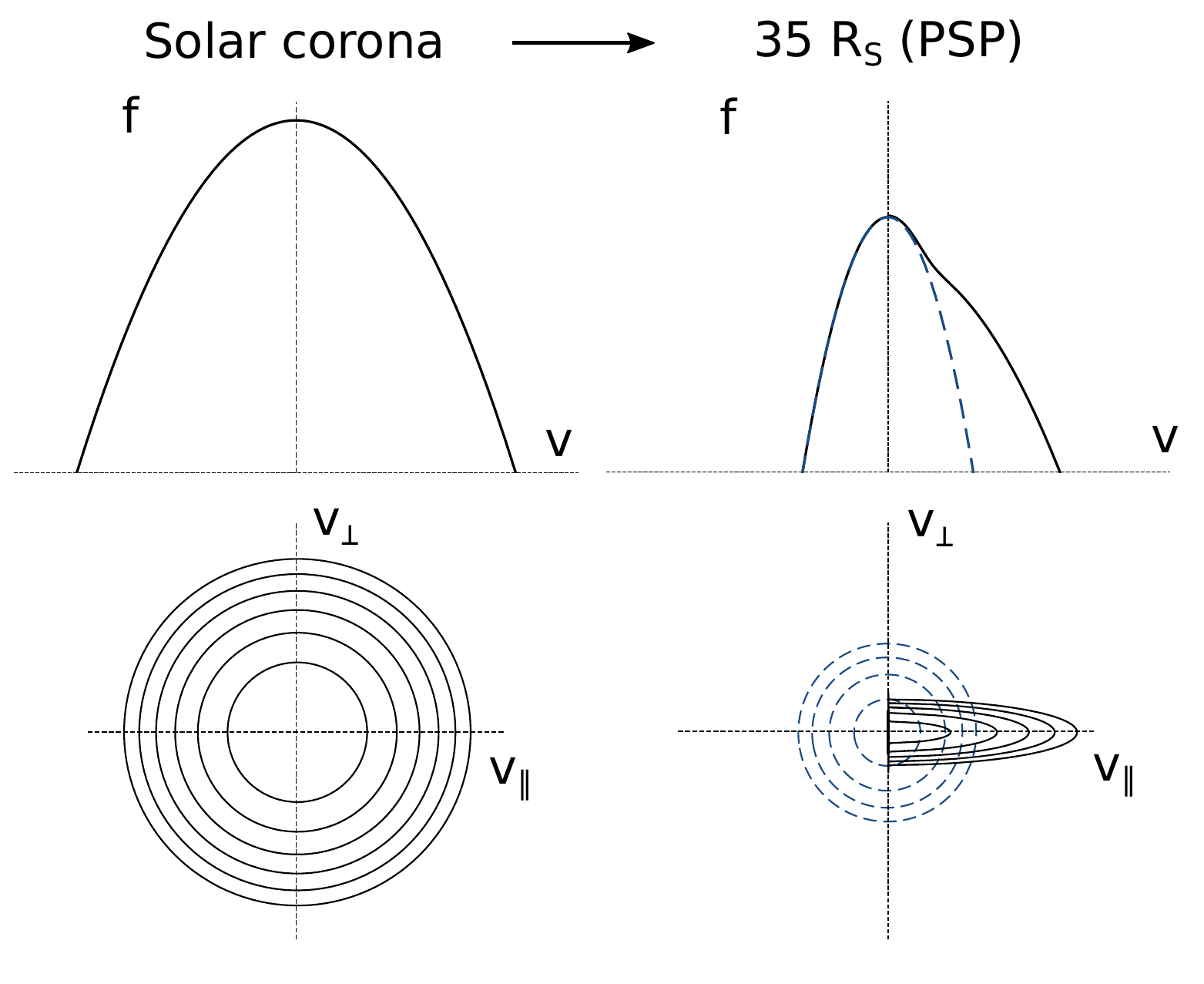}
\caption{A schematics demonstrating how the information about the temperature of coronal electrons is preserved in the T$_{s\parallel}$. The upper row shows a cut through a distribution function with respect to electron velocity, while the lower row shows the same two distribution in velocity space where v$_\parallel$ is aligned with magnetic field direction.}
\label{fig:schematics}%
\end{figure}

In reality the transition between collision dominated and collisionless regime does not happen at one distance, but is a continuous process. This could be one of the reasons why our single exobase focusing model predicts lower PAWs than observed. Another possibility is that the strahl has already been affected by scattering mechanisms also resulting in an anti-correlation between PAW and electron energy. A good candidate are Coulomb collisions. A study using kinetic theory is presented in works by \citet{Horaites2018a, Horaites2019}, providing a theoretical prediction of the strahl PAW, accounting for collisions between particles. PAW seems, to some extent and for some energy range, to agree with the results from \citet{Horaites2018a}, predicting relations in the form PAW $\propto$ $\sqrt{n}$, and PAW $\propto$ E$^{-1}$, where n strands for density and E for electron energy. However, Eq. 15 from \citet{Horaites2019} does not predict well our observations. \\

The focusing experienced by the strahl electron component during the solar wind expansion does not affect T$_{s\parallel}$. If the scattering mechanisms do not strongly modify the electron VDF, the temperature of the VDF at the exobase stays imprinted in the strahl population. The density of the VDFs and the core T decrease with radial distance, but the slope - the temperature - of the parallel cut through the strahl part of the VDF remains unchanged (see schematics in Fig. \ref{fig:schematics}). Even though the strahl PAW are observed to be somewhat broader than predicted by the exospheric models, no radial trends were found in the T$_{s\parallel}$ observed by Helios and PSP missions (see Figs. \ref{fig:tsm-rad-helios} and \ref{fig:tsm-rad-psp}). This raises the question about how efficient are the scattering mechanisms in modifying T$_{s\parallel}$. The answer requires further observational and numerical studies, out of the scope of the present work.

\begin{figure}
\centering
\includegraphics[width=1\hsize]{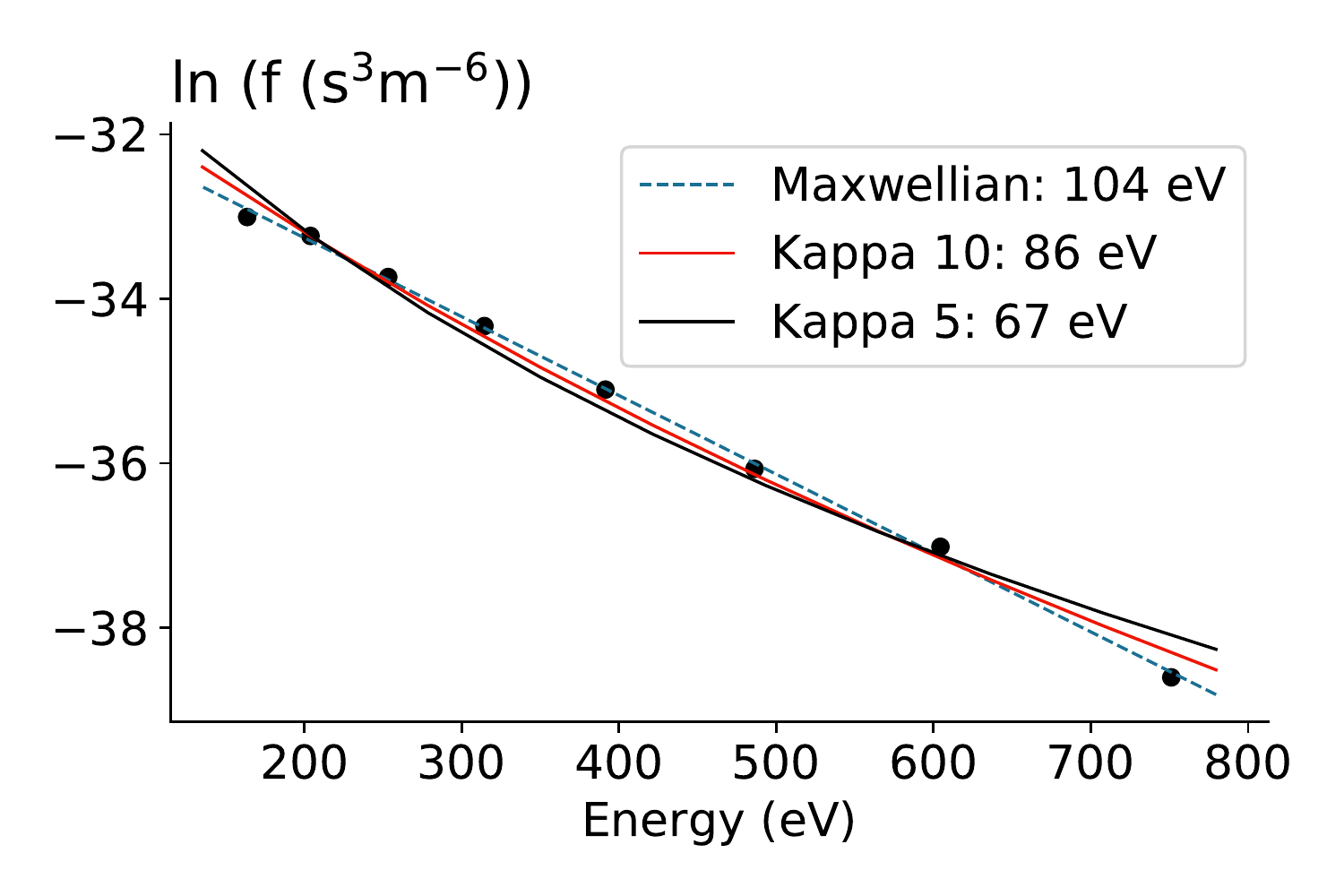}
\caption{The same example as shown in Fig. \ref{fig:pawfit} (c). Additionally to the Maxwellian fit (dashed blue line) two Kappa fits are shown: $\kappa$ = 10 (red line) and $\kappa$ = 5 (black line).}
\label{fig:maxkap}%
\end{figure}

The constant behaviour of T$_{s\parallel}$ over radial distance is a new observation, which is in contradiction with the current beliefs about its radial evolution. Recent near-Earth observational studies, from either Wind or Cluster spacecrafts \citep{vinas2012,Tao2016,MacNeil2017,Wilson2019b}, report significantly lower strahl T to the ones reported in this work. As a consequence it was accepted that the strahl temperature must decrease over the distance from Sun to Earth. But the reason for different observational results lies mostly in the different data analysis approach. We follow the exospheric theory and therefore fit the strahl field-aligned VDF cut with a Maxwellian centred on $v_\parallel$ = 0. \citet{vinas2012} isolate the strahl component and obtain the plasma moments by integration. The obtained strahl T are on the order of 10 eV with T$_\perp$ $\sim$ 2 times bigger than T$_\parallel$, and the obtained drift velocities are relatively high. With this approach they measure the width of the strahl distribution, while we are interested in the slope. \citet{Wilson2019b} model the strahl with a Kappa distribution with a drift and report the mean T$_{s\parallel, \kappa}$ of 44.2 eV. \citet{Tao2016} and \citet{MacNeil2017} also use Kappa distribution function, but centred on $v_\parallel$ = 0, and find the mean T$_{s\parallel, \kappa}$ of 51.1 eV and $\sim$ 50 eV, respectively. We performed a test to quantify the effect of the different model choice on the obtained T$_\parallel$. Fig. \ref{fig:maxkap} shows the same example as Fig. \ref{fig:pawfit}(c) but fitted with a Kappa distribution function for $\kappa$ = 10 and $\kappa$ = 5. The fit was made in the logarithmic space with the 1-dimensional Kappa distribution function:

\begin{equation}
f_\kappa (v_\parallel) = \frac{n_\kappa}{\kappa^{3/2} \sqrt{\pi} w_\kappa} \frac{\Gamma(\kappa+1)}{\Gamma(\kappa-1/2)} \cdot \Big( 1 + \frac{v_\parallel^2}{w_\kappa^2 \kappa} \Big)^{-\kappa-1},
\label{eq:kappa}
\end{equation}

where $\kappa$ parameter is given, and the density ($n_\kappa$) and the thermal velocity ($w_\kappa$) are the fitting parameters. The strahl parallel Kappa temperature (T$_{s\parallel, \kappa}$) can be calculated from $w_\kappa$:

\begin{equation}
w_\kappa = \sqrt{\frac{2 \kappa -3}{\kappa k_B T_{s\parallel, \kappa}}}.
\label{eq:kappa_w}
\end{equation}

In fact, the temperature obtained with a Kappa fit (T$_{s\parallel, \kappa}$) is much lower than the temperature obtained with a Maxwellian fit. The T$_{s\parallel, \kappa}$ for $\kappa$ = 5 case falls within the range of observations shown by \citet{Tao2016} and \citet{MacNeil2017}.

A Maxwellian model was chosen because it most accurately represents new observations of the strahl provided by PSP. For the measured energy range, up to 800 eV, the strahl VDF cut through the parallel direction shows no signs of high energy tails. This is not strictly true for the strahl measured by Helios, especially at larger distances, neither for the strahl observed at 1 au \citep{Tao2016, MacNeil2017, Wilson2019b}. The radial evolution of the strahl modelled by a Kappa function is presented by \citet{Stverak2009a}. They report the increase of $\kappa$ values at smaller distances from the Sun, reaching $\sim$ 14 at 0.3 au (64.5 R$_S$). For large $\kappa$ values a Kappa distribution tends towards a Maxwellian, therefore, the PSP observations of a Maxwellian-like strahl VDF below 60 R$_S$ were not unexpected.

On the other hand, the increase of $\kappa$ for small distances \citep{Stverak2009a}, and the measured Maxwellian-like strahls by the PSP, could be an instrumental artefact. Closer to the Sun the total electron temperature is larger, therefore the unaltered instrument energy span becomes relatively smaller than further from the Sun. We are only able to measure a smaller portion of the strahl VDF, and could be missing the high energy tails, the most important part for distinction between a Kappa and a Maxwellian VDF.
\\

The idea that the strahl electrons carry the information about the temperature of the electrons at the exobase is a part of exospheric solar wind models as they initially do not take into account collisions, or wave-particle interactions \citep{Jockers1970,LemaireJosephandScherer1971}. Two studies investigated the relation between the temperature of the supra-thermal electron population observed at 1 au, and the coronal temperature so far. Both of them make use of the oxygen charge state ratio (O$^{7+}$/O$^{6+}$) measurements as an estimate for the coronal electron temperature. In the study by \citet{Hefti2008} a clear correlation between the strahl parallel temperature and the oxygen charge ratio was observed, while \citet{MacNeil2017} present a data interval where the correlation is present and a data interval where it is not. Unfortunately neither of the two spacecraft presented in this work provide us with the measurement of oxygen state ratios. But this measurement will be provided by the Solar Orbiter, the new probe exploring the inner heliosphere launched on Feb 10 2020 \citep{Muller2013}.

Another parameter strongly correlated with the temperature of the solar corona which is often used as an indicator of the solar wind origin is the solar wind velocity \citep{lopezfreeman1986}. The solar wind originating from the centre of the coronal holes, where the proton plasma temperature is higher than that of electrons, has higher terminal velocity than the wind coming from the edges of the coronal holes. Figs. \ref{fig:tsm-rad-helios} and \ref{fig:tsm-rad-psp} agree with this global picture as they display a clear anti-correlation between T$_{s\parallel}$ and the solar wind proton velocity. \\

Using the sub-spacecraft points in combination with the synoptic map of the Sun allows us to follow the time evolution of the T$_{s\parallel}$ and compare it with current state of solar corona (Fig. \ref{fig:tsm-sub-psp}). Through comparison with the PFSS modelling of the magnetic field line topology during the first orbit of PSP presented in the study by \citet{Badman2019}, we crudely separated the strahl temperature data into 3 bins. The coldest T$_{s\parallel}$ (T$_{s\parallel}$ $<$ 75 eV) were observed at times when measured magnetic field lines appear to connect to a bigger equatorial coronal hole encountered just after the first PSP perihelion. During the first encounter when a period of high-speed solar wind implies connectivity to the smaller coronal hole (Fig. \ref{fig:tsm-sub-psp} (b)) the strahl temperatures appear a bit higher temperature, 75 eV $<$ T$_{s\parallel}$ $<$ 85 eV. These values are in agreement with the coronal electron temperatures obtained through spectroscopy technique presented by \citet{David1998, Cranmer2002}. They report coronal hole electron temperature just above the solar surface to be 0.79 MK (= 68 eV), reach the maximum temperature at 1.15 R$_S$ and stays below 1 MK (= 86 eV) and decreases after (Fig. 2 in the referred article). For the quiet equatorial corona the temperatures appear to be higher, starting at 1 MK and increasing reaching 3.16 MK (= 272 eV) at 1.3 R$_S$.  

This evidence lead us to believe that T$_{s\parallel}$ indeed retains the information about the temperature of electrons at their origin. However, to be convinced that T$_{s\parallel}$ is not just correlated with, but equals to the coronal electron temperature further analysis is required. As mentioned above, the exobase is not a discrete point above the solar surface but a continuous region over which the collisions become less and less important. Another thing one needs to account for is the energy dependant scattering of the strahl electrons. The strahl was for most of the measurements observed to be narrow, but still broader than what is expected from the simple collisionsless model. For example, scattering by Coulomb collisions at only lower energies would result in a higher T$_{s\parallel}$. The study of the effect of continuous exobase and Coulomb collisions making use of kinetic simulations BiCop \citep{LandiPantellini2001, Landi2003a} is a current work in progress.

\section{Conclusions}

The PAW data obtained during the first two orbits of PSP agrees well with the results obtained from Helios data set presented by \citet{Bercic2019}. We find the same PAW dependence on $\beta_{ec\parallel}$: in high-$\beta_{ec\parallel}$ solar wind the strahl appears broader than in the low-$\beta_{ec\parallel}$ solar wind. For the measured energy range the PAW was found to decrease with electron energy reaching down to 30$^o$ for the lower two $\beta_{ec\parallel}$ bins representing the majority of measurements. \\

We present for the first time observations of T$_{s\parallel}$ from both, PSP and Helios missions. An anti-correlation was found between T$_{s\parallel}$ and the solar wind velocity (v$_p$), while T$_{s\parallel}$ was observed to be constant over radial distance (r). These findings lead us to conclude that the strahl carries the information about the coronal electron temperature at the point of its origin and can be used as a good proxy for the connectivity studies involving remote sensing and in-situ data. In fact, the origins of the solar wind measured by PSP anticipated from the strahl temperature measurements compare very well to the ones obtained using a PFSS model presented by \citet{Bale2019, Badman2019}. Even though the measured values of T$_{s\parallel}$ agree very well with the coronal electron temperatures measured with the spectroscopes on-board SOHO spacecraft \citep{David1998, Cranmer2002}, we believe further analysis is required to confirm that T$_{s\parallel}$ is a direct measure of the electron temperature in corona.

\FloatBarrier

\begin{appendix}
\section{Pitch-angle width - FOV effects}
\label{app:fov}
The combined FOV of SPAN-E instruments is not uniform \citep{Kasper2016, Whittlesey2020}. Electrons are detected by two instruments and the azimuth anodes of each of them have two possible angular widths, 6$^o$ and 24$^o$. Part of the full solid angle is not sampled, and part blocked by the heat shield (see Fig. \ref{fig:fov}). We investigated how much the nonuniform FOV affects our data analysis.  

We identified two extreme configurations of the magnetic field vector in the FOV. The measurement is the most precise when the magnetic field vector lies within part of the FOV covered by the small azimuth anodes of SPAN-A. The most problematic measurement of the strahl electrons happens when the magnetic field is aligned with the radial direction, because in this case the strahl electrons get blocked by the heat shield which is during the encounter time directed directly towards the Sun.

\begin{figure*}
\centering
\includegraphics[width=1\hsize]{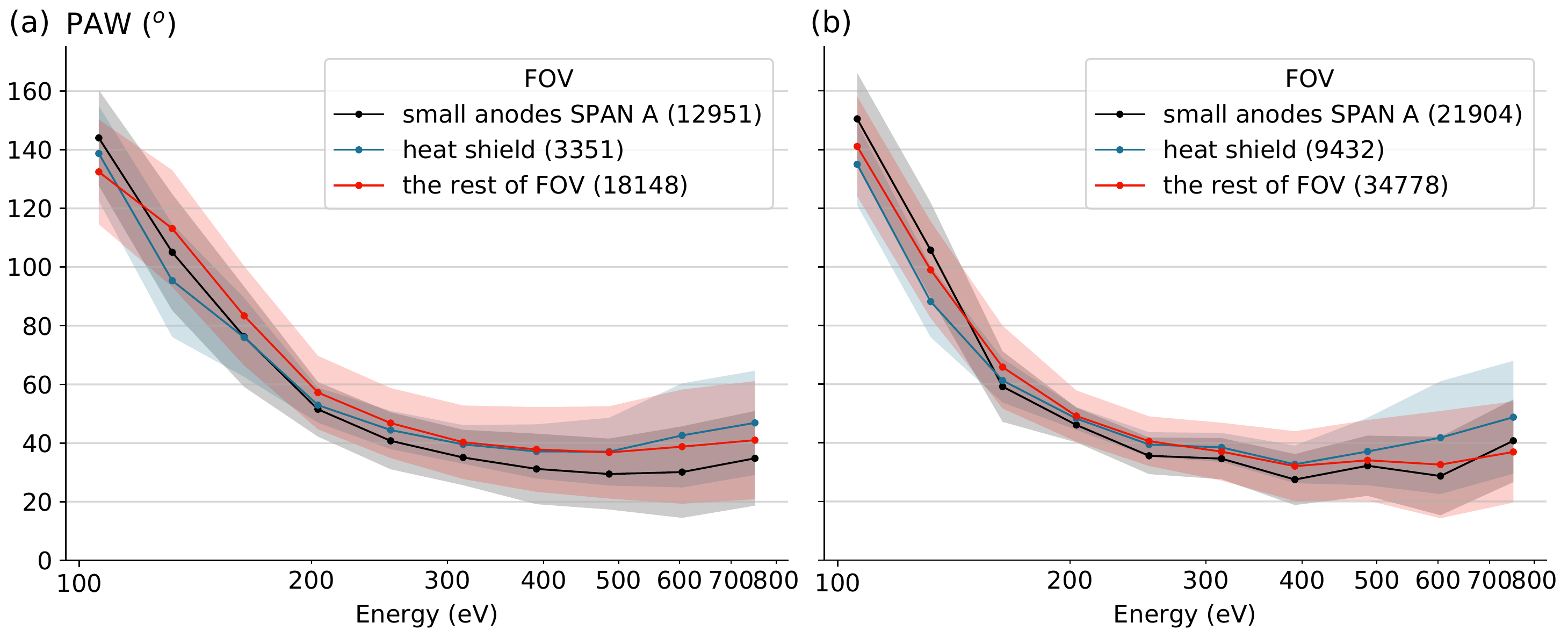}
\caption{Strahl PAW with respect to electron energy, separated into three bins according to location of the magnetic field vector in the FOV of the instruments. Bin edges are noted in the legend followed by a number of instances belonging to each bin. (a) Encounter 1, (b) Encounter 2.}
\label{fig:paw-fov}%
\end{figure*}

The results are presented in Fig. \ref{fig:paw-fov}. PAWs measured at low electron energies are independent of configuration of the magnetic field in the FOV. However, the strahl electrons with higher energies during the first encounter appear $\sim$ 10$^o$ broader when the magnetic field lies outside of the FOV covered by the small anodes of SPAN-A. The variation is less pronounced during the second encounter.

\section{Pitch-angle width - standard deviation B}
\label{app:Bstd}
Another test was conducted to determine how much the variation of the magnetic field during the integration time of SPAN-E instruments affects our strahl PAW measurements. We calculated the standard deviation of a 294 Hz magnetic field measurement during each SPAN-E scan lasting 27.96 s for the first encounter. For the second encounter magnetic field was sampled with a cadence of 147 Hz and the duration of one scan set to 13.98 s.

\begin{figure*}
\centering
\includegraphics[width=1\hsize]{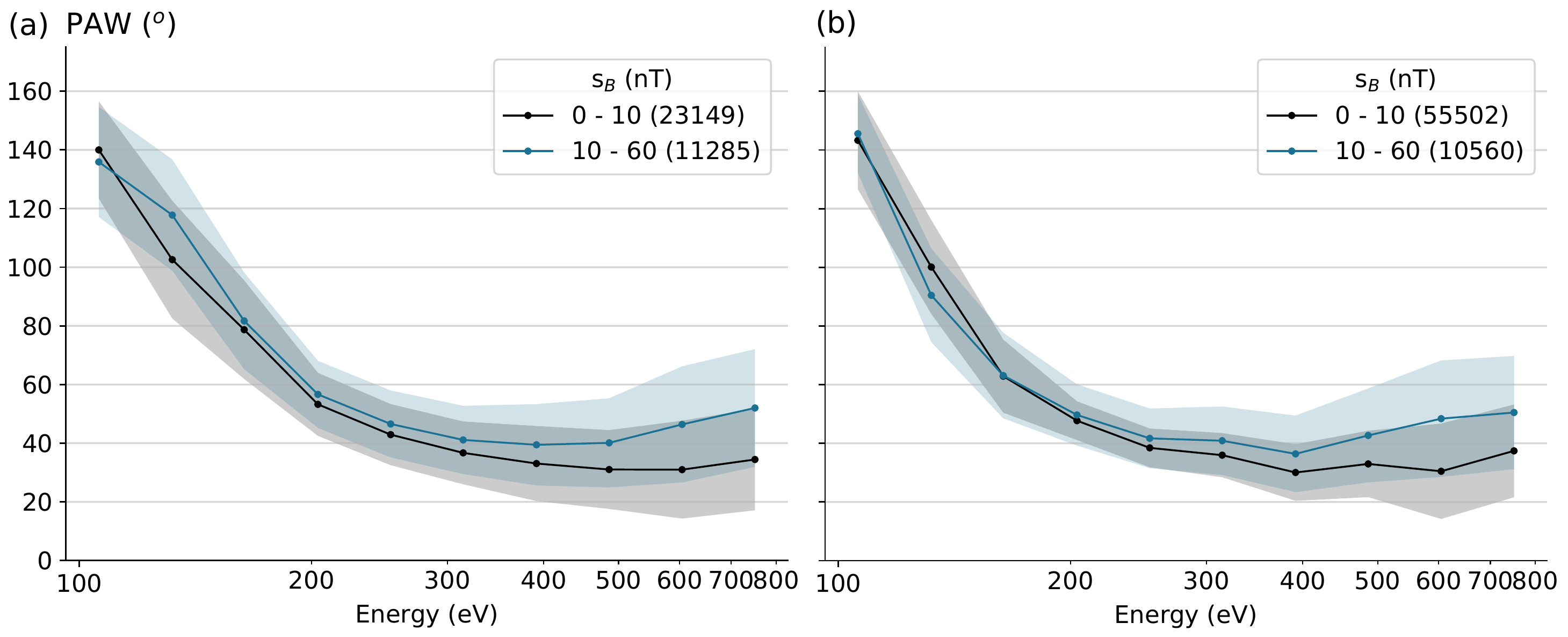}
\caption{Strahl PAW with respect to electron energy, separated into two bins according to the standard deviation of the magnetic field measurement during the duration of one scan.  (a) Encounter 1, (b) Encounter 2.}
\label{fig:paw-magstd}%
\end{figure*}

Fig. \ref{fig:paw-magstd} shows PAWs, separated into two groups according to the standard deviation of B. The effect of the strong variation of the magnetic field during SPAN-E measurement is increasing with increasing electron energy and can make the strahl appear up to 20$^o$ broader than during times of small magnetic field variation.

\end{appendix}

\section*{Acknowledgements}
We are grateful for the reviewers comments which were constructive and helped to improve the quality of the present work. This work was supported by the Programme National PNST of CNRS/INSU co-funded by CNES. All the analysis was done, and the plots produced using open source Python libraries NumPy, Matplotlib, Pandas, and SciPy. We also acknowledge all members of the Helios data archive team (\url{http://helios-data.ssl.berkeley.edu/team-members/}) to make the Helios data publicly available to the space physics community. 

\bibliography{psp_strahl}{}
\bibliographystyle{aasjournal}



\end{document}